\documentclass{PoS}

\usepackage[dvipsnames]{xcolor}

\def\ttl#1{{\it #1}}
\def\floatcaption#1#2{ \caption{#2 \label{#1}} }
\def\hf{{\hat{f}}}
\def\hB{{\hat{B}}}
\def\hM{{\hat{M}}}
\def\tr{{\rm tr}}

\title{The large-mass regime of confining but nearly conformal gauge theories}

\ShortTitle{The large-mass regime}

\author{\speaker{Maarten Golterman}\\
        Department of Physics and Astronomy, San Francisco State University,\\
        San Francisco, CA 94132, USA\\
        and\\
        Department of Physics and IFAE-BIST, Universitat Aut\`onoma de Barcelona,\\
        E-08193 Bellaterra, Barcelona, Spain\\
        E-mail: \email{maarten@sfsu.edu}}

\author{Yigal Shamir\\
        Raymond and Beverly Sackler School of Physics and Astronomy,
        Tel~Aviv University,\\ 69978, Tel~Aviv, Israel\\
        E-mail: \email{shamir@post.tau.ac.il}}

\abstract{We apply a recently developed dilaton-pion effective field theory for
asymptotically free gauge theories near the conformal window to the $SU(3)$
gauge theory with $N_f=8$ fermions in the fundamental representation.
Numerical data for this theory suggest the existence of a large-mass
regime, where the fermion mass is not small but nevertheless the effective
theory is applicable because of the parametric proximity of the conformal
window.  In this regime, we find that the mass dependence of hadronic
quantities is similar to that of a a mass-deformed conformal theory, so
that distinguishing infrared conformality from confinement requires the
study of subleading effects.}

\FullConference{The 36th Annual International Symposium on Lattice Field Theory - LATTICE2018\\
		22-28 July, 2018\\
		Michigan State University, East Lansing, Michigan, USA.}

\begin{document}

\section{Introduction}
Triggered by numerical results \cite{Appelquist:2016viq,Aoki:2016wnc,Fodor:2017nlp,Appelquist:2018yqe}\footnote{The three latter references quoted here represent the most
recent numerical work.  For more extensive earlier references, see Ref.~\cite{Golterman:2016lsd}.} obtained in simulations of gauge theories with fermions
which are generally believed to be near the conformal window, we developed a systematic
effective field theory (EFT) to describe the long-distance behavior of such theories, specifically,
those with an $SU(N_c)$ gauge group with $N_f$ fermions in the fundamental representation
\cite{Golterman:2016lsd}.   In this talk, we report on more recent work aiming to compare the
EFT directly with numerical data \cite{Golterman:2018mfm}.   In particular, Ref.~\cite{Golterman:2018mfm} made the following claims:
\begin{itemize}
\item
Even if the fermion mass of a near-conformal theory is large
compared to the chiral-symmetry breaking/confinement scale,
it can be described systematically by the EFT.
\item
In this large-mass regime, it is very hard to distinguish such a
theory from a mass-perturbed theory whose massless limit is conformal in the
infrared.
\item
In the case of the $SU(3)$, $N_f = 8$ theory, either much smaller
fermion masses or a much higher precision will be needed in
current simulations.
\end{itemize}
The rest of this talk will be concerned with a brief review of the explanations for these
claims.

\section{The effective field theory}

Figure~\ref{LSD} shows the hadron spectrum for $SU(3)$ gauge theory with $N_f=8$ degenerate fundamental flavors, as a function of the fermion mass $m$.  Quite unlike the more familiar case
of QCD, the spectrum contains a very light, and, for these fermion masses, stable $0^{++}$
state, nearly degenerate with the pseudo-Nambu--Goldstone bosons for the $SU(8)_L\times
SU(8)_R$ chiral symmetry breaking, which we will refer to as the pions.

Clearly, any EFT aiming to describe the low-energy pion physics should also contain a field
for the $0^{++}$ state, which, in light of what follows, we will refer to as the dilatonic meson
(dilaton, for short).
In order to construct such a theory, we made the following assumptions \cite{Golterman:2016lsd}:
\begin{itemize}
\item
Scale invariance gets restored as we take the theory closer to the
conformal window. For $N_f$ fundamental flavors in an $SU(N_c)$ theory
this happens when $N_f$ crosses into the conformal region.
Technically, $n_f\equiv N_f/N_c$ attains the critical value $n_f^*$ at this crossing
in the limit $N_c$, $N_f\to\infty$.
\item
The theory contains pions associated with chiral symmetry breaking,
and a dilaton associated with breaking of scale symmetry, which
becomes massless for $n_f\to n_f^*$ (and $m\to 0$).
\item
The dilaton potential has a zero, as a function of the dilaton field $\tau$.
\end{itemize}
Under these assumptions, the following leading-order (LO) lagrangian was
constructed in Refs.~\cite{Golterman:2016lsd,Golterman:2018mfm}:
\begin{eqnarray}
\label{lag}
{\cal L}&=&\frac{\hf_\pi^2}{4}\,e^{2\tau}\tr(\partial_\mu\Sigma^\dagger\partial_\mu\Sigma)
+\frac{\hf_\tau^2}{2}\,e^{2\tau}(\partial_\mu\tau)^2\\
&&-\frac{\hf_\pi^2\hB_\pi m}{2}\,e^{(3-\gamma_*)\tau}\,\tr(\Sigma+\Sigma^\dagger)
+\hf_\tau^2\hB_\tau\,e^{4\tau}\,c_1\left(\tau-\frac{1}{4}\right)\ .\nonumber
\end{eqnarray}
Here $\gamma_*$ is the value of mass anomalous dimension at the infrared fixed point
at $n_f=n_f^*$.   Along with the mass $m$, the parameter $c_1\propto n_f-n_f^*$ is
assumed to be small, and the EFT allows a systematic expansion in these small
parameters.  The $\tau$ field has been shifted so that its LO expectation value $v\equiv\langle\tau\rangle=0$ when $m=0$.

\begin{figure}[t]
\vspace*{4ex}
\begin{center}
\includegraphics*[width=8cm,angle=-90,origin=c]{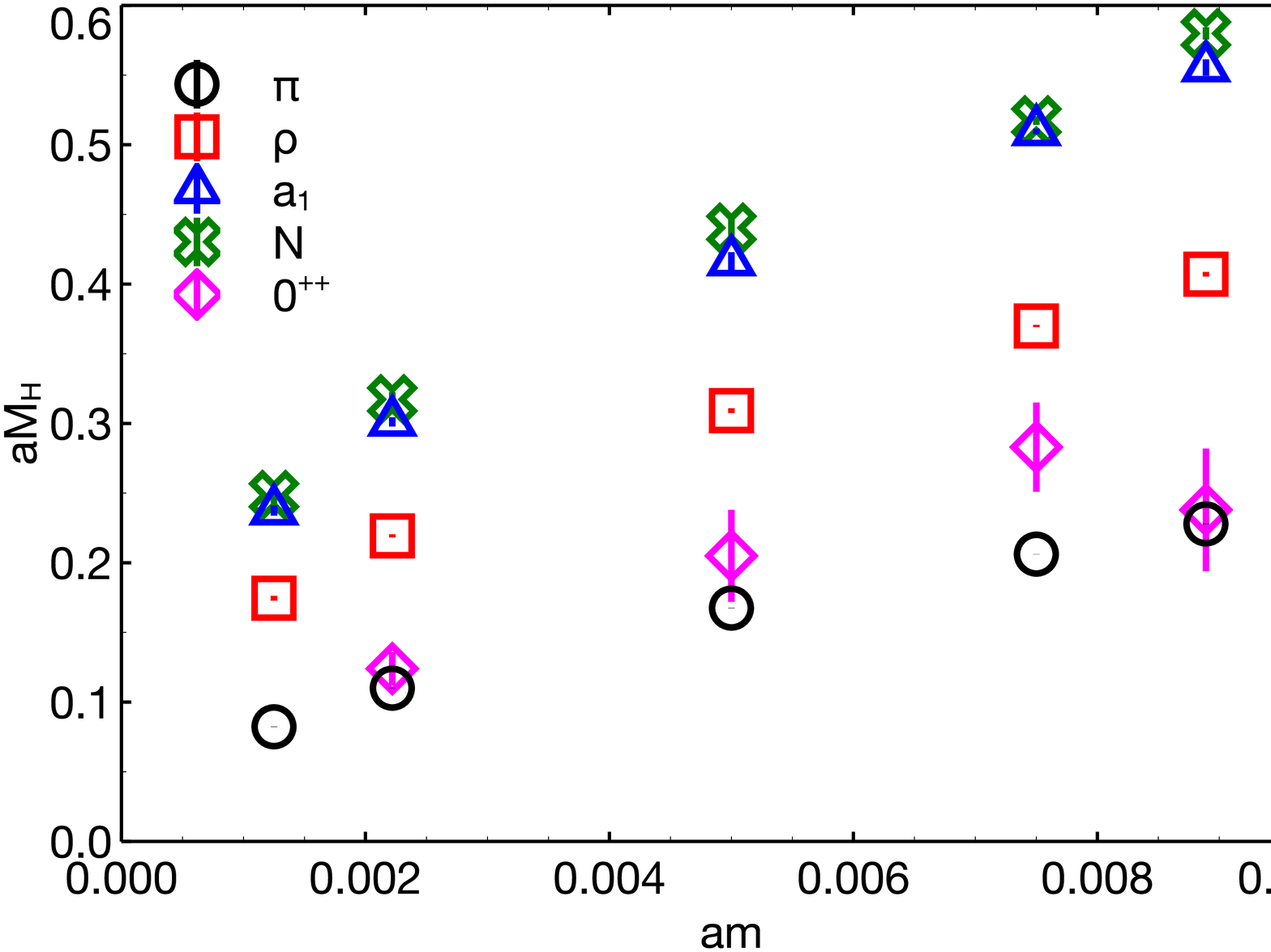}
\end{center}
\vspace*{-8ex}
\begin{quotation}
\vspace*{-4ex}
\floatcaption{LSD}%
{{\it The hadron spectrum for $SU(3)$ gauge theory with $N_f=8$ degenerate
fundamental flavors, from Ref.~{\rm \cite{Appelquist:2016viq}}.}}
\end{quotation}
\end{figure}

\section{Lowest-order predictions}
At non-vanishing $m$, the LO value for $v$ satisfies the equation
\begin{equation}
\label{mintree}
\frac{m}{c_1\hM}=v\,e^{(1+\gamma_*)v}\ ,
\end{equation}
where
\begin{equation}
\label{Mhat}
\hM=\frac{4\hf_\tau^2\hB_\tau}{\hf_\pi^2\hB_\pi N_f(3-\gamma_3)}\ .
\end{equation}
We note that this is an ``order-1'' equation: both $m$ and $c_1$ are small, and
we are assuming a power counting in which they are parametrically of the same
size.

If $m$ is large enough that the left-hand side of Eq.~(\ref{mintree}) is much larger
than one (we will refer to this range as the ``large-mass regime''), we find the approximate solution
\begin{equation}
\label{approx}
e^v\approx \left(\frac{m}{c_1\hM}\right)^{1/(1+\gamma_*)}\ .
\end{equation}
From this, it follows that {\it all} masses and decay constants scale as $m^{1/(1+\gamma_*)}$.
In other words, even though the theory is in the phase in which chiral symmetry is broken
spontaneously at $m=0$, all masses and decay constants exhibit hyperscaling behavior as
in a mass-perturbed conformal theory \cite{DelDebbio:2010hx}!

In more detail, let us see how this arises for a few examples.   Using exact LO results for
physical quantities, and then relation~(\ref{approx}), we find, for instance,
\begin{eqnarray}
\label{examples}
F_\pi&=&\hf_\pi\,e^v\propto\hf_\pi\left(\frac{m}{c_1}\right)^{1/(1+\gamma_*)}\ ,\\
M_\pi^2&=&2\hB_\pi m\,e^{(1-\gamma_*)v}\propto 2\hB_\pi\,c_1 \left(\frac{m}{c_1}\right)^{2/(1+\gamma_*)}\ ,\nonumber\\
M_{\rm nucleon}&=&M_0\,e^v\propto M_0 \left(\frac{m}{c_1}\right)^{1/(1+\gamma_*)}\ ,\nonumber
\end{eqnarray}
where $M_0$ is the nucleon mass in the chiral limit.\footnote{Note that for small $m$, the solution~(\ref{approx}) is not valid, and the nucleon mass does not vanish in the chiral limit,
as long as $n_f<n_f^*$.}  All expressions on the right-hand side of these equations show
the claimed large-mass hyperscaling behavior.

These results are valid as long as loop corrections are small.   As usual, loop corrections
are governed by loop expansion parameters such as $M_\pi^2/(4\pi F_\pi)^2$ and
$M_\tau^2/(4\pi F_\tau)^2$.   Considering the first of these two, we find using the exact
LO relations in Eq.~(\ref{examples}) and Eq.~(\ref{mintree}), that
\begin{equation}
\label{loop}
\frac{M_\pi^2}{(4\pi F_\pi)^2}=\frac{2c_1\hM\hB_\pi}{(4\pi\hf_\pi)^2}\,v\ .
\end{equation}
This is small as long as $c_1\log{m}$ is small, where we used that, in the large-mass
regime, $v\sim\log{m}$.   This establishes the range of validity of the EFT; for masses
larger than this value, loop corrections become too large for the loop
expansion defined by the EFT to be useful.

\section{Comparison with numerical results}

From our result, that in the large-mass regime all masses and decay constants
exhibit hyperscaling, it follows that ratios of masses and decay constants are
approximately constant in this regime.   Indeed, results obtained in
Ref.~\cite{Appelquist:2016viq}, shown in Fig.~\ref{ratios},
show approximate constancy of the ratio of
five different hadron masses with the pion decay constant, as a function of $m$.

Of course, we need to check whether our LO analysis applies to these data.   Indeed,
the value of the expansion parameter~(\ref{loop}) is roughly equal to $0.1$ for the ratios
shown in Fig.~\ref{ratios}.   However, as also discussed in Ref.~\cite{Appelquist:2017vyy}, it is
possible that loop corrections to some quantities are larger than suggested by this value.

The consistency between our hyperscaling prediction and the data strongly suggests that the values of $m$ investigated in Ref.~\cite{Appelquist:2018yqe}
are in the large-mass regime.   And indeed, this turns out to be the case.   First, fitting the
quantity $M_\pi^2 F_\pi^{-1+\gamma_*}=2\hB_\pi\hf^{-1+\gamma_*}m$ as a function
of $m$ yields the estimate $\gamma_*\approx 1$ \cite{Appelquist:2017vyy}.
Using this, and values for the nucleon mass as a function of $m$ obtained in
Ref.~\cite{Appelquist:2018yqe}, we find that at the lightest value of $m$ \cite{Golterman:2018mfm},
\begin{equation}
\label{massvalue}
am=0.00125\qquad\Rightarrow\qquad\frac{m}{c_1\hM}\sim 100\ .
\end{equation}
This implies that, to unambiguously determine whether this theory is conformal
or breaks chiral-symmetry in the massless limit, we need either much smaller masses, or enough
precision to disentangle subleading effects to the large-mass behavior.

As an additional point of interest, we also find from the numerical data that
$\hB_\pi/\hf_\pi\sim 10^3$, which may be a sign of ``condensate enhancement''
in this theory.

\begin{figure}[t]
\vspace*{4ex}
\begin{center}
\includegraphics*[width=11cm]{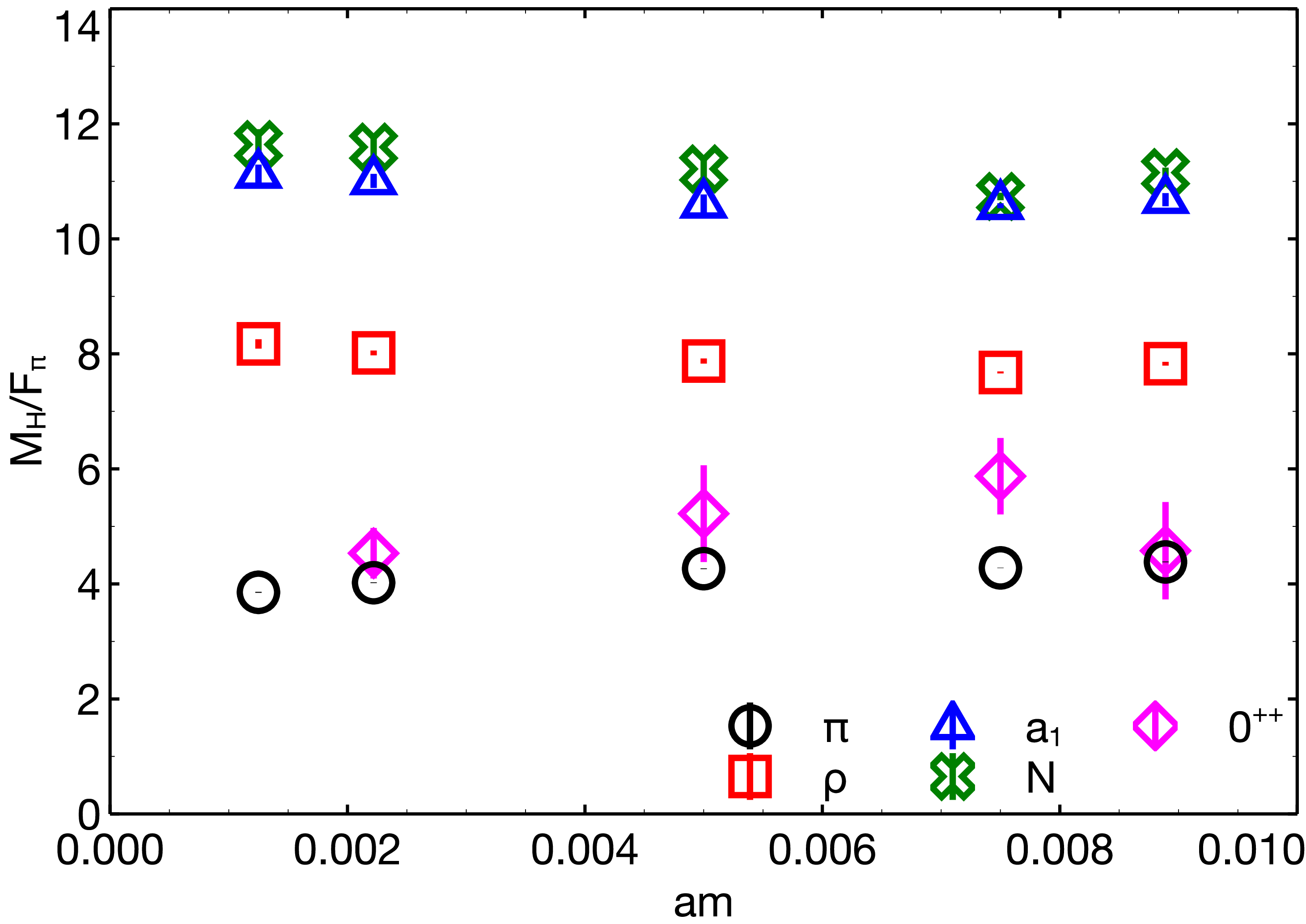}
\end{center}
\begin{quotation}
\floatcaption{ratios}%
{{\it The hadron spectrum for $SU(3)$ gauge theory with $N_f=8$ degenerate
fundamental flavors, from Ref.~{\rm \cite{Appelquist:2016viq}}.}}
\end{quotation}
\end{figure}

\section{Conclusion}
We briefly summarize our conclusions, referring to Ref.~\cite{Golterman:2018mfm} for
much more detail.

First, we find that in near-conformal theories, there exists a ``large-mass'' regime, defined
by $m/(c_1\hM)\gg 1$, while the region of applicability of the EFT is, parametrically,
$c_1\log{m}\ll 1$.   In this regime, masses and decay constants show approximate
hyperscaling like in a mass-deformed conformal theory, even though the theory
breaks chiral symmetry at $m=0$.

Second, we find that current simulations in the $SU(3)$, $N_f=8$ theory are in the
large-mass regime, with $m/(c_1\hM)\ge 100$.   Therefore, in this theory, either much
smaller masses or much higher precision will be needed to disentangle subleading
effects, and distinguish, numerically, between a mass-deformed conformal theory
and a theory with spontaneous chiral symmetry breaking in the massless limit.

Summarizing, assuming the validity of our EFT, it provides a possible explanation
why it is so difficult to distinguish a conformal theory from a chiral-symmetry
breaking theory near the conformal sill.

\section*{Acknowledgements}
This material is based upon work supported by the U.S. Department of
Energy, Office of Science, Office of High Energy Physics, under Award
Number DE-FG03-92ER40711 (MG).
YS is supported by the Israel Science Foundation
under grant no.~491/17.

\end{document}